\documentclass[reprint,aip,jcp,amsmath,citeautoscript]{revtex4-1}

\usepackage{graphicx}
\usepackage{braket}
\usepackage{txfonts}
\usepackage{verbatim}
\usepackage{color}
\definecolor{myblue}{rgb}{0,0,1}
\usepackage[breaklinks=true,colorlinks=true,linkcolor=myblue,urlcolor=myblue,citecolor=myblue]{hyperref}

\newcommand{\opk}[1]{{\hat{#1}}}
\newcommand{\clk}[1]{{#1}}

\newcommand{\varN}{{30}}
\newcommand{\varJ}{{1.0}}
\newcommand{\vargsq}{{5.0}}
\newcommand{\varomega}{{0.1}}
\newcommand{\varT}{{1.0}}
\newcommand{\vardt}{{0.02}}
\newcommand{\varsamp}{{12\;000}}

\begin{document}

\title{A Reciprocal-Space Formulation of Mixed Quantum-Classical Dynamics}

\author{Alex Krotz}
\author{Justin Provazza}
\author{Roel Tempelaar}
\email{roel.tempelaar@northwestern.edu}

\affiliation{Department of Chemistry, Northwestern University, 2145 Sheridan Road, Evanston, Illinois 60208, USA}

\begin{abstract}
We derive a formulation of mixed quantum-classical dynamics for describing electronic carriers interacting with phonons in reciprocal space. For dispersionless phonons, we start by expressing the real-space classical coordinates in terms of complex variables. A Fourier series over these coordinates then yields the reciprocal-space coordinates. Evaluating the electron-phonon interaction term through Ehrenfest's theorem, we arrive at a reciprocal-space formalism that is equivalent to mean-field mixed quantum-classical dynamics in real space. This equivalence is numerically verified for the Holstein and Peierls models, for which we find the reciprocal-space Hellmann--Feynman forces to involve momentum derivative contributions in addition to the position derivative terms commonly seen in real space. We close by presenting a proof of concept for the inexpensive modeling of low-momentum carriers interacting with phonons by means of a truncated basis in reciprocal space, which is not possible within a real space formulation.
\end{abstract}

\maketitle

\emph{Introduction.} Real space and reciprocal space provide two alternative representations for describing quantum-mechanical phenomena in the condensed phase. Typically, finite-sized disordered solids such as molecules and molecular aggregates are characterized by localized quantum excitations and carrier dynamics dominated by an incoherent hopping between sites. Both aspects are described most effectively in real space. For (quasi)infinite and periodic solids such as highly-ordered crystals, on the other hand, quantum states take the form of Bloch waves, and carrier dynamics proceeds through ``bandlike'' transport, for which a reciprocal space representation is the most effective.

Materials for which reciprocal space representations have traditionally been adopted are typically characterized by a weak coupling between electronic carriers and nuclear vibrations (phonons). This coupling, which underpins the nonequilibrium dynamics of such materials, could therefore be adequately represented by theories truncating higher-order electron-phonon correlations \cite{lindberg1988effective}. This is markedly different for molecular materials, where electron-phonon coupling is intermediate to strong. The need to accurately model interacting electrons and phonons in such materials has resulted in a plethora of quantum-dynamical methods, among which those involving an expansion and/or projection of the electron-phonon interaction terms \cite{nakajima1958quantum, zwanzig1960ensemble, tanimura1989time, makri1995numerical, shi2003new, segal2010numerically, huo2011partial, liu2014reduced, chen2015dynamics, dunn2019removing, pfalzgraff2019efficient, yan2020new, mulvihill2020modified}, polaron transforms \cite{fetherolf2020unification}, and tensor-network decompositions \cite{worth1996effect, wang2003multilayer, prior2010efficient, kurashige2018matrix, kloss2019multiset}, each conventionally represented in real space. These methods in principle retain a quantum treatment for all explicit degrees of freedom, which comes with high computational cost, especially when interactions are treated nonperturbatively. An inexpensive alternative is provided by mixed quantum-classical methods where phonons are treated classically, which enables one to describe nonperturbative and Markovian dynamics at the expense of taking the classical approximation. Mixed quantum-classical dynamics has proven successful in describing a broad variety of molecular phenomena \cite{subotnik20016understanding, crespo-otero2018recent, nelson2020nonadiabatic}.

The recent years have seen an increased interest in materials that are best represented in reciprocal space, but which feature intermediate to strong electron-phonon coupling. Examples of such materials include monolayer and few-layer variants of transition-metal dichalcogenides \cite{mak2010atomically, splendiani2010emerging, shree2018observation, trovatello2020strongly, li2021exciton} as well as layered and bulk hybrid metal-halide perovskites \cite{kojima2009organometal, zhu2015charge, wright2016electron}, both of which have risen to prominence due to their potential application as optoelectronic and quantum information devices. The rational engineering of such materials for technological purposes relies on a thorough understanding of their nonequilibrium properties, which requires the development of reciprocal-space models accounting for electron-phonon interactions beyond a truncated/perturbative level \cite{mayers2018how, lengers2020theory, brem2020phonon}.

Here, we derive a mixed quantum-classical formalism tailored to interacting electronic carriers and phonons in reciprocal space. Assuming the local sites to form a periodic lattice, and the phonon modes to be dispersionless and harmonic, we first express the real-space classical phonon coordinates in terms of complex variables, which when taken as a Fourier series yields classical expressions in reciprocal space. These expressions are shown to be identical to those obtained by a Fourier transform of the quantum Hamiltonian followed by the classical approximation applied in reciprocal space. We evaluate the transformed electron-phonon interaction term by invoking Ehrenfest's theorem \cite{ehrenfest1927bemerkung}, effectively describing the Hellmann-Feynman forces acting on the classical modes by means of a mean-field average of the quantum state \cite{mclachlan1964variational, micha1983self, kirson1984dynamics, sawada1985mean, berendsen1993quantum}. Whereas in real space these forces commonly involve a position derivative term, an additional momentum derivative contribution is shown to appear in reciprocal space, which contributes to the classical equations of motion. We apply the resulting approach to the Holstein and Peierls models, both of which offer a straightforward comparison between our approach and the conventional real-space formulation of mean-field mixed quantum-classical dynamics, verifying the equivalence between the two formalisms. For both
models, we present a proof of concept for the inexpensive modeling of low-momentum carriers interacting with phonons by means of a truncation of the Brillouin zone, which can only be realized in a reciprocal-space formulation.

\emph{Classical phonons in reciprocal space.} For a one-dimensional lattice of harmonic, noninteracting, and dispersionless (Einstein) phonons the real-space quantum Hamiltonian is given by
\begin{align}
    \hat{H}_\text{ph}
    =\omega\sum_n\Big( \hat{b}_n^\dagger \hat{b}_n+\frac{1}{2}\Big)
    =\sum_n\Big(\frac{1}{2}\hat{p}_n^2+\frac{1}{2}\omega^2\hat{q}_n^2\Big),
    \label{eq_H_ph}
\end{align}
where $\omega$ is the phonon energy ($\hbar=1$ is taken here and throughout), and $\hat{q}_n$, $\hat{p}_n$, and $\hat{b}_n^{(\dagger)}$ are the (mass-weighted) position, momentum, and ladder operators for the local mode at lattice site $n$. Applying the classical approximation to this Hamiltonian amounts to replacing the position and momentum operators by their classical coordinates,
\begin{align}
    \hat{q}_n \rightarrow q_n,\quad \hat{p}_n \rightarrow p_n.
\end{align}
Introducing the classical equivalent of the ladder operators,
\begin{align}
    z_n\equiv\sqrt{\frac{\omega}{2}}\bigg(q_n+i\frac{p_n}{\omega}\bigg),
\end{align}
the classical Hamiltonian can be expressed as
\footnote{Note that the zero-point contribution does not show up as a result of the classical position and momentum operators commuting, as expected.}
\begin{align}
    H_\text{ph}=\sum_n\bigg(\frac{1}{2}p_n^2+\frac{1}{2}\omega^2 q_n^2\bigg)=\omega\sum_n z_n^*z_n.
\end{align}
Using Hamilton's equations for $q_n$ and $p_n$ it can be shown that the time derivative of $z_n$ is given by $\dot{z}_n=-i\omega z_n$.

Similarly to the ladder operators, a Fourier series over $z_n$ yields its reciprocal-space equivalent as
\begin{align}
    \clk{z}_k=\frac{1}{\sqrt{N}}\sum_n e^{ikn}z_n.
\end{align}
Here, the lattice is assumed to be periodic and to consist of $N$ sites, and $k$ is the wavevector in units of $2\pi/a$, with $a$ as the lattice constant.
(Note that in the following we will consistently use $n$ to denote sites, and $k$ and $\kappa$ to denote wavevectors.) The classical Hamiltonian can then be expressed in reciprocal space as
\begin{align}
    \clk{H}_\text{ph}=\omega\sum_k \clk{z}_k^*\clk{z}_k=\sum_k\Big(\frac{1}{2}\clk{p}_k^2+\frac{1}{2}\omega^2 \clk{q}_k^2\Big),
    \label{eq_H_ph_classical_k}
\end{align}
with the reciprocal ``position'' and ``momentum'' coordinates given by
\begin{align}
    \clk{q}_k&=\frac{1}{\sqrt{2\omega}}(\clk{z}_k+\clk{z}_k^*)
    =\frac{1}{\sqrt{N}}\sum_n\bigg(q_n\cos(kn)-\frac{p_n}{\omega}\sin(kn)\bigg)\nonumber,
    \\
    \clk{p}_k&=-i\sqrt{\frac{\omega}{2}}(\clk{z}_k-\clk{z}_k^*)
    =\frac{\omega}{\sqrt{N}}\sum_n\bigg(\frac{p_n}{\omega}\cos(kn)+q_n\sin(kn)\bigg).
    \label{eq_qk_pk}
\end{align}
It is straightforward to show that $\dot{\clk{z}}_k=-i\omega\clk{z}_k$, and that the time-evolution of $\clk{q}_k$ and $\clk{p}_k$ is governed by Hamilton's equations using $\clk{H}_\text{ph}$ given by Eq.~\ref{eq_H_ph_classical_k}. The result is formally equivalent to the real-space equations of motion, but is expressed entirely within reciprocal space.

It is worth noting that Eq.~\ref{eq_H_ph_classical_k} is identical to the result obtained when first Fourier-transforming the quantum Hamiltonian to reciprocal space,
\begin{align}
    \opk{H}_\text{ph}
    =\omega\sum_k\Big( \opk{b}_k^\dagger \opk{b}_k+\frac{1}{2}\Big)
    =\sum_k\Big(\frac{1}{2}\opk{p}_k^2+\frac{1}{2}\omega^2\opk{q}_k^2\Big),
    \label{eq_H_ph_quantum_k}
\end{align}
and then taking the classical approximation by replacing
$\opk{q}_k \rightarrow \clk{q}_k$ and $\opk{p}_k \rightarrow \clk{p}_k$. This observation is a manifestation of the equivalence of canonical representations in classical-limit quantum mechanics \cite{miller1974classical}.

\emph{Mixed quantum-classical system.} In the following, we consider a system of interacting electronic carriers and phonons on a periodic lattice. The total quantum Hamiltonian is partitioned as
\begin{align}
    \hat{H}=\hat{H}_\text{el}+\hat{H}_\text{ph}+\hat{H}_\text{el-ph}.
\end{align}
with $\hat{H}_\text{ph}$ given by Eq.~\ref{eq_H_ph} or \ref{eq_H_ph_quantum_k}. For the electronic part, we consider a simple tight-binding model given by the real-space Hamiltonian
\begin{align}
    \hat{H}_\text{el}=-J\sum_n\left(\hat{c}^{\dagger}_{n+1}\hat{c}_{n} + \hat{c}^{\dagger}_{n}\hat{c}_{n+1}\right),
\end{align}
where $\hat{c}_n^{(\dagger)}$ are the ladder operators for an electronic carrier at site $n$, and where $J$ is the nearest-neighbor interaction term. (Note that due to periodic boundaries, site $N+1$ couples to site $1$.) The reciprocal analog is given by
\begin{align}
    \opk{H}_\text{el}=-2J\sum_k\opk{c}_k^\dagger\opk{c}_k\cos(k).
\end{align}

Within the classical approximation for the phonons $\hat{H}_\text{ph}=H_\text{ph}$, and the electron-phonon quantum Hamiltonian $\hat{H}_\text{el-ph}$ depends parametrically on the phonon coordinates. Not only do the phonon coordinates impact the electronic quantum states through this parametric dependence, they also experience a ``quantum force'' due to the electronic states, which affects their classical equations of motion. The phonon momentum term is offdiagonal in real space, as a result of which only the phonon position contributes to $\hat{H}_\text{el-ph}$. The quantum force acting on the local phonon mode at site $n$ then follows from the Hellmann--Feynman theorem as
\begin{align}
    F_n=-\braket{\Psi|\nabla_{q_n}\hat{H}_\text{el-ph}|\Psi}.
\end{align}
Here, $\Psi$ is the electronic state that provides ``feedback'' to the phonon coordinates. Mixed quantum-classical methods vary in their choice of feedback state \cite{crespo-otero2018recent}. In the present study we restrict ourselves to mean-field mixed quantum-classical dynamics \cite{mclachlan1964variational, micha1983self, kirson1984dynamics, sawada1985mean, berendsen1993quantum}, where $\Psi$ is chosen to be an electronic superposition state that is propagated by the Schr\"odinger equation, 
\begin{align}
    \dot{\Psi}=-i(\hat{H}_\text{el}+\hat{H}_\text{el-ph})\Psi.
\end{align}
This choice of feedback state is motivated by Ehrenfest's theorem stating that the classical analog of a quantum state behaves as the quantum expectation value with respect to that state \cite{ehrenfest1927bemerkung}. 

From Eq.~\ref{eq_qk_pk} it can be seen that the classical position and momentum coordinates become scrambled under the transformation to their reciprocal analogs. As a result, the Hellmann--Feynman forces appearing in the reciprocal-space equations of motion involve both position and momentum derivative terms, with Hamilton's equations given by
\begin{align}
    \dot{\clk{p}}_k&=-\frac{\partial (\clk{H}_\text{ph}+\clk{H}_\text{el-ph})}{\partial \clk{q}_k}
    =-\omega^2\clk{q}_k-\braket{\Psi|\nabla_{\clk{q}_k}\opk{H}_\text{el-ph}|\Psi}\nonumber,\\
    \dot{\clk{q}}_k&=\frac{\partial (\clk{H}_\text{ph}+\clk{H}_\text{el-ph})}{\partial \clk{p}_k}
    =\clk{p}_k+\braket{\Psi|\nabla_{\clk{p}_k}\opk{H}_\text{el-ph}|\Psi}.
\end{align}

\emph{Holstein and Peierls models.} Using the above approach, we proceed by considering the Holstein and Peierls models, which differ by the form of the electron-phonon interaction Hamiltonian. For the Holstein model, the real-space phonon position coordinates couple linearly and diagonally to the local quantum states with the electron-phonon Hamiltonian expressed in terms of quantum operators as
\begin{align}
    \hat{H}_\text{el-ph}&=g\omega \sum_{n}\hat{c}_{n}^{\dagger}\hat{c}_{n}\left(\hat{b}_{n}^{\dagger} + \hat{b}_{n}\right)
    =g\sqrt{2\omega^{3}}\sum_{n}\hat{c}_{n}^{\dagger}\hat{c}_{n}\hat{q}_{n}.
\end{align}
Here, $g$ is the dimensionless coupling parameter, which relates to the vibrational reorganization energy as $g^2\omega$ \footnote{Strictly speaking, this energy amount needs to be added to the diagonal entries of the total Hamiltonian $\hat{H}$ for each electronic particle. This has been omitted in the present work since it does not change the dynamics within the single-particle manifold.}. The reciprocal-space Hamiltonian is given by
\begin{align}
    \opk{H}_\text{el-ph}&=\frac{g\omega}{\sqrt{N}}\sum_{k,\kappa}\opk{c}_{k+\kappa}^\dagger \opk{c}_k \left(\opk{b}^\dagger_{-\kappa}+\opk{b}_\kappa\right)\nonumber\\
    &=\frac{g\sqrt{\omega}}{\sqrt{2N}}\sum_{k,\kappa}\opk{c}_{k+\kappa}^\dagger \opk{c}_k
    \Big(\omega\big(\opk{q}_{-\kappa}+\opk{q}_\kappa\big)-i\big(\opk{p}_{-\kappa}-\opk{p}_\kappa\big)\Big).
\end{align}
Taking the classical approximation (either in real space or reciprocal space) yields the position and momentum derivative terms appearing in the reciprocal-space Hamilton's equations
\begin{align}
    \braket{\Psi|\nabla_{\clk{q}_k}\opk{H}_\text{el-ph}|\Psi}&=g\sqrt{\frac{2\omega^3}{N}}\Re\{C_k\},
    \nonumber\\
    \braket{\Psi|\nabla_{\clk{p}_k}\opk{H}_\text{el-ph}|\Psi}&=-g\sqrt{\frac{2\omega}{N}}\Im\{C_k\},
\end{align}
with the autocorrelation function
\begin{align}
    C_k\equiv\sum_{k'}\braket{\Psi|\opk{c}_{k'+k}^\dagger \opk{c}_{k'}|\Psi}.
\end{align}
Note that while $\opk{H}_\text{el-ph}$ is complex-valued, the position and momentum derivative terms are purely real.

For the Peierls model (also known as Su--Schrieffer--Heeger model \cite{su1979solitons}), the real-space phonon position coordinates couple linearly to the electronic nearest-neighbor interaction terms as
\begin{align}
    \hat{H}_\text{el-ph}
    &=g\omega\sum_{n}\left(\hat{c}^{\dagger}_{n}\hat{c}_{n+1}+\hat{c}^{\dagger}_{n+1}\hat{c}_{n}\right)
    \left(\hat{b}^{\dagger}_{n}+\hat{b}_{n}-\hat{b}^{\dagger}_{n+1}-\hat{b}_{n+1}\right)\nonumber\\
    &=g\sqrt{2\omega^{3}}\sum_{n}\left(\hat{c}^{\dagger}_{n}\hat{c}_{n+1}+\hat{c}^{\dagger}_{n+1}\hat{c}_{n}\right)\left(\hat{q}_{n}-\hat{q}_{n+1}\right),
\end{align}
which in reciprocal space yields
\begin{align}
    \opk{H}_\text{el-ph}&=2i\frac{g\omega}{\sqrt{N}}\sum_{k,\kappa}\opk{c}_{k+\kappa}^\dagger \opk{c}_k \Big(\opk{b}^\dagger_{-\kappa}+\opk{b}_\kappa\Big)
    \big(\sin(k+\kappa)-\sin(k)\big)\nonumber\\
    &=\frac{g\sqrt{2\omega}}{\sqrt{N}}\sum_{k,\kappa}\opk{c}_{k+\kappa}^\dagger \opk{c}_k \left(i\omega\big(\opk{q}_{-\kappa}+\opk{q}_\kappa\big)+\big(\opk{p}_{-\kappa}-\opk{p}_\kappa\big)\right)\nonumber\\
    &\qquad\times\left(\sin(k+\kappa)-\sin(k)\right),
\end{align}
where $g$ again denotes the dimensionless coupling parameter. This yields the position and momentum derivative terms
\begin{align}
    \braket{\Psi|\nabla_{\clk{q}_k}\opk{H}_\text{el-ph}|\Psi}&=-g\sqrt{\frac{2\omega^3}{N}}\Im\left\{C_k'\right\},\nonumber\\
    \braket{\Psi|\nabla_{\clk{p}_k}\opk{H}_\text{el-ph}|\Psi}&=-g\sqrt{\frac{2\omega}{N}}\Re\left\{C_k'\right\},
\end{align}
with the modulated autocorrelation function
\begin{align}
    C_k'\equiv\sum_{k'}\braket{\Psi|\opk{c}_{k'+k}^\dagger \opk{c}_{k'}|\Psi}\big(\sin(k'+k)-\sin(k')\big).
\end{align}

As seen above, within the Holstein and Peierls models we find the mixed quantum-classical equations of motion to assume simple forms in both real space and reciprocal space. It is therefore straightforward to numerically verify the equivalence between the real-space and reciprocal-space formulations. In doing so, we restrict ourselves to the single-quantum manifold of $\hat{H}_\text{e}$ (considering a single electronic carrier), with the initial electronic state taken to be the $k=0$ state,
\begin{align}
    \ket{\Psi}=\ket{k=0}=\frac{1}{\sqrt{N}}\sum_n\ket{n}.
    \label{eq_el_initial}
\end{align}
Such an initial condition could be representative of a tightly-bound electron-hole pair (Frenkel exciton) created upon impulsive optical excitation \footnote{An optically excited exciton involves a negligible wavevector since the Fermi sea has $k=0$ and the absorbed photon momentum is negligible.}, in which case $n$ represents the exciton location. The initial values of the classical coordinates $q_n$ and $p_n$ are sampled independently from a Boltzmann distribution,
\begin{align}
    P\big(\{q_n,p_n\}\big) \propto \prod_n \exp\left(-\beta\frac{1}{2}\left(p^{2}_{n}+\omega^{2} q^{2}_{n}\right)
    \right),
\end{align}
where $\beta=1/T$ is the inverse temperature ($k_\text{B}=1$ is taken). Through a Fourier transform of this expression one finds an identical distribution for the reciprocal-space coordinates $\clk{q}_k$ and $\clk{p}_k$.

\begin{figure}
\includegraphics[scale=1.0]{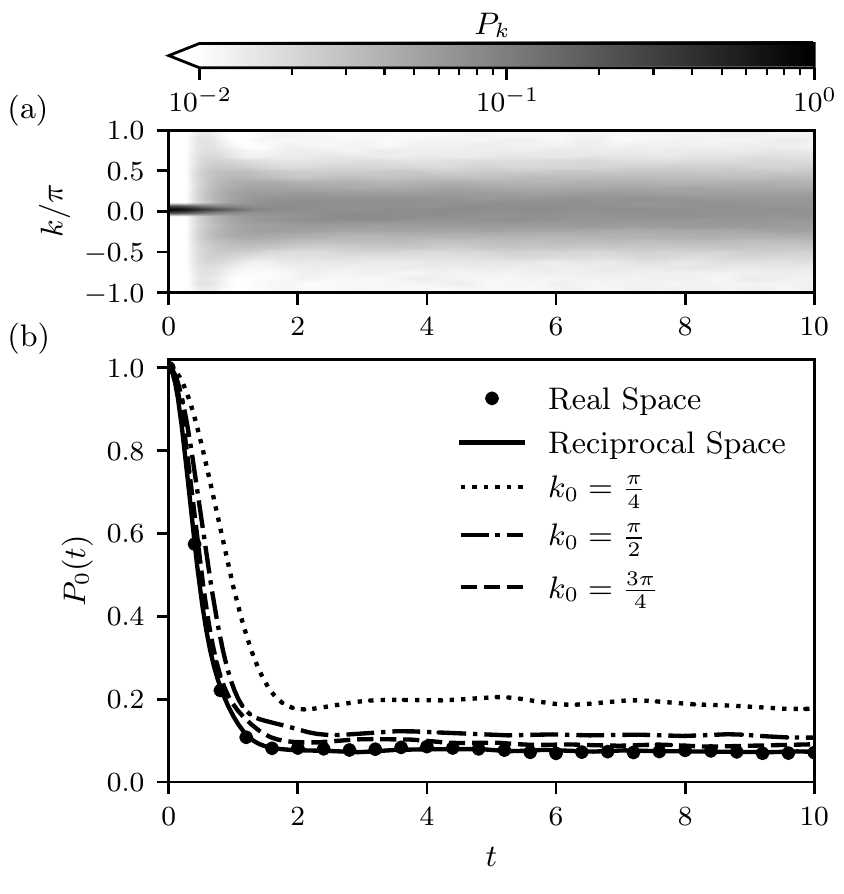}
\caption{Transient electronic populations $P_k(t)$ calculated within the mean-field mixed quantum-classical method for the Holstein model with $J=\varJ$, $\omega=\varomega$, $g^2=\vargsq$, and $T=\varT$. Shown in (a) are $k$-dependent populations obtained through the reciprocal-space formulation of the method. Shown in (b) is $P_0(t)$ obtained through the real-space (markers) and reciprocal-space (solid) formalisms. Also shown are reciprocal-space results obtained upon truncating the Brillouin beyond $\vert k\vert<k_0$.}
\label{fig:holstein}
\end{figure}

Fig.~\ref{fig:holstein} shows numerical results for a periodic lattice consisting of $N=\varN$ sites and with $J=\varJ$, $\omega=\varomega$, $g^2=\vargsq$, and $T=\varT$ within the Holstein model. Parameters are expressed without units to keep the discussion general, but we note that when taking the thermal energy at room temperature ($T=293$~K) as a reference, a unit of energy amounts to 25~meV and a unit of time to 164~fs. Results were obtained by propagating the classical and quantum coordinates using a fourth-order Runge-Kutta algorithm with a time step of $\Delta t=\vardt$. All results have been averaged over $\varsamp$ thermal initial conditions for the classical coordinates. Shown in Fig.~\ref{fig:holstein} (a) are time-dependent reciprocal-space populations of the electronic carrier $P_k(t)\equiv\vert\braket{k|\Psi(t)}\vert^2$, calculated by solving the equations of motion for all coordinates in reciprocal space. The electronic state, upon initiating at $k=0$, can be seen to rapidly broaden in reciprocal space due to scattering with the phonon modes, while equilibrating within a time span of $t\sim2$. Throughout, the populations are seen to remain symmetric with respect to inversion of $k$, as a result of the underlying Hamiltonians and (thermally-averaged) initial conditions being conserved under this symmetry operation. Also notable is that the majority of the populations remain concentrated at $k=0$. The reason for this behavior is that the electronic band is minimized at this Brillouin zone location, and that the relatively large number of phonon modes forces the electronic carrier to relax to a quasi-thermal equilibrium, which favors low energies. It should be pointed out that mean-field mixed quantum-classical dynamics is known violate detailed balance, but nevertheless induces a low-energy bias \cite{tully1998mixed, parandekar2005mixed}.

Fig.~\ref{fig:holstein} (b) compares the time-dependent populations at $k=0$, $P_0(t)$, resulting from the reciprocal-space and real-space mixed quantum-classical formalisms. (The local populations coming out of the real-space formulation have been Fourier-transformed in order to yield $P_0(t)$.) The lack of conceivable differences between the shown data confirms the equivalence between the equations of motion in both formalisms. We found this result to be independent on the choice of parameters.

\begin{figure}
\includegraphics[scale=1.0]{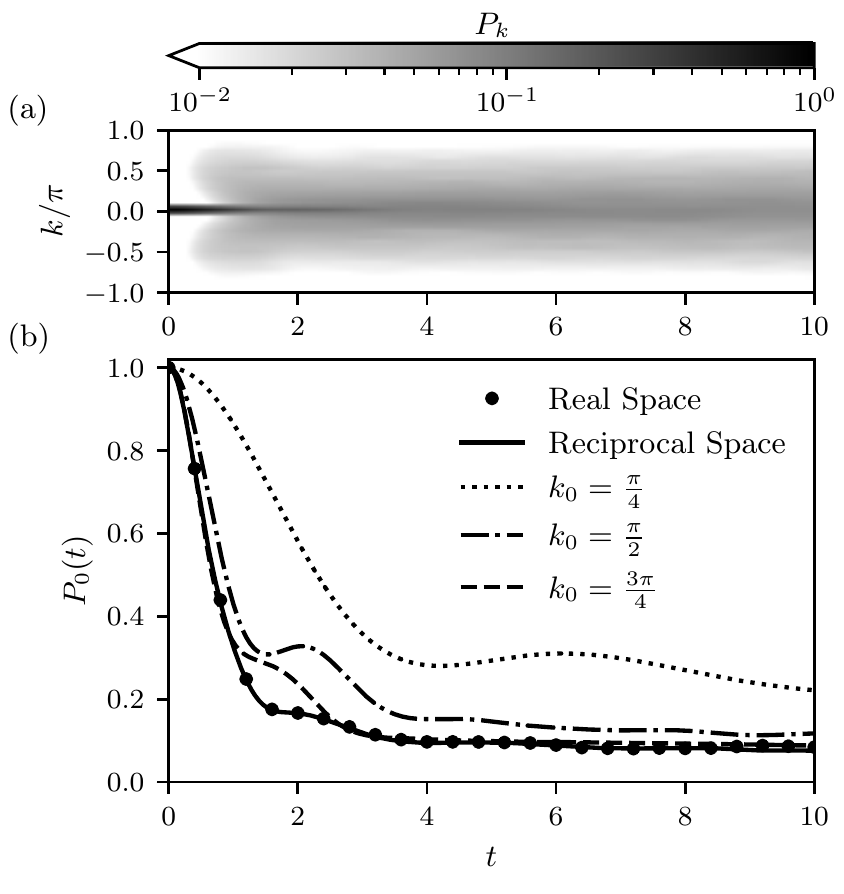}
\caption{Same as Fig.~\ref{fig:holstein} but for the Peierls model.}
\label{fig:peierls}
\end{figure}

Shown in Fig.~\ref{fig:peierls} are results analogous to Fig.~\ref{fig:holstein}, but for the Peierls model. Again, a rapid broadening in reciprocal space is observed. Similarly to the Holstein model, the formal equivalence between the real-space and reciprocal-space mixed quantum-classical equations of motion yields a perfect agreement between results obtained in both representations.

\emph{Brillouin zone truncation.} The appeal of a real-space representation is that local phenomena can be accurately described at manageable computational cost by truncating the degrees of freedom beyond those within the spatial domain of interest. A similar truncation can be applied to reciprocal space for phenomena that occur within a limited domain of the Brillouin zone. Electronic carriers may remain confined to a limited domain when the time scales under consideration do not allow them to cover the full Brillouin zone, or when they thermally relax to band minima. This principle has been been utilized in static calculations of exciton \cite{qiu2016screening}
and trion \cite{tempelaar2019many} states in monolayer transition-metal dichalcogenides, where a truncation radius around the $K$ points in the Brillouin zone was imposed. To the best of our knowledge, applications of similar reciprocal-space truncation schemes in dynamical calculations have remained limited.

From the results in Figs.~\ref{fig:holstein} (a) and \ref{fig:peierls} (a) it can be seen that the electronic populations continuously remain concentrated around $k=0$ for both Holstein and Peierls models, owing to the $k=0$ initial condition and as well as the band minimum being located here. This suggests that accurate results are retained by restricting the Brillouin zone to within a truncation radius, $k_0$, such that the electronic basis states and phonon modes are limited to those having $\vert k\vert<k_0$. To demonstrate that this is indeed the case, we show in Figs.~\ref{fig:holstein} (b) and \ref{fig:peierls} (b) results for various values of $k_0$. As can be seen, for the Holstein model accurate dynamics is obtained even upon halving the Brillouin zone. For a similar truncation within the Peierls model, discrepancies can be seen to emerge around $t\approx2$, but the short and long-time dynamics are still reasonably accurate. Needless to say, such truncated results could only be obtained using the reciprocal-space mixed quantum-classical method, as any truncation in real space would grossly distort even the initial state which is delocalized over the entire lattice -- see Eq.~\ref{eq_el_initial}.

\emph{Discussion and conclusions.} While mixed quantum-classical methods are conventionally formulated in real space, we have shown here that for a periodic lattice an equivalent reciprocal-space formulation can be obtained by taking a Fourier series over the real-space classical coordinates expressed in terms of complex variables. As such, we have arrived at an approach tailored to describing bandlike phenomena such as electronic carriers interacting with phonons in crystalline solids. Some of the benefits offered by a reciprocal-space representation are illustrated in the present study by its proof of concept for the accurate modeling of low-momentum carriers using a truncated Brillouin zone.

We have specifically considered a mean-field approach to self-consistently describe the electron-phonon interactions, in which case we find the real-space and reciprocal-space mixed quantum-classical formalisms to be equivalent. In verifying this equivalence through numerical calculations for the Holstein and Peierls models, we have chosen the parameters to be such that the intermediate coupling regime is reached (where perturbative approaches loose validity), and that the classical approximation should hold reasonably well ($T\gg\omega$). It is important to reiterate, however, that the real-space and reciprocal-space results come out identically regardless of the choice of parameters.

While it is tempting to assess the accuracy of the dynamics obtained in this work against exact results, it is important to note that the reciprocal-space formulation is merely a convenient representation of a method that has been well-characterized in real space \cite{tully1998mixed, parandekar2005mixed, crespo-otero2018recent, xie2020performance}, and as such its accuracy should be no different than that of the real-space variant. In particular, mean-field mixed quantum-classical dynamics has previously been shown to yield inaccurate equilibrium populations as a result of its violation of detailed balance \cite{tully1998mixed, parandekar2005mixed}. Indeed, for the parameters employed in the present work we found the quantum populations to equilibrate towards an effective temperature significantly exceeding $T$. It would therefore be worthwhile to consider alternative formulations for the electron-photon interaction term known to yield an improved description of detailed balance \cite{bastida2006modified}. Interestingly, this would considerably enhance the long-time accuracy for the truncated data shown in Fig.~\ref{fig:peierls} (b), as a decrease in the effective temperature will localize the quantum populations to a narrower domain around $k=0$. It is also noteworthy that the numerical efficiency for obtaining accurate results upon Brillouin zone truncations is expected to increase with higher dimensionality, for which a generalization of our formalism is trivial.

Upon obtaining the reciprocal-space mixed quantum-classical equations of motion through a Fourier transform of the real-space classical phonon coordinates, we observed that identical equations would be obtained upon taking the classical approximation within reciprocal space (i.e., upon first Fourier-transforming the quantum phonon coordinates). This opens the opportunity to perform mixed quantum-classical modeling of reciprocal space Hamiltonians that have no simple analog in real space, such as those having dispersed phonons. More broadly speaking, it would be interesting to explore the ideas presented in the present work in the context of anharmonic modes and nonlinear electron-phonon coupling. It is conceivable that some of the many benefits of classical molecular dynamics can be harnessed while taking advantage of a reciprocal-space representation. 

\bibliography{Bibliography}

\end{document}